\documentstyle[12pt]{article}
\begin{document}
\newcommand{\be}{\begin{equation}}
\newcommand{\ee}{\end{equation}}
\newcommand{\bea}{\begin{eqnarray}}
\newcommand{\eea}{\end{eqnarray}}
\newcommand{\beas}{\begin{eqnarray*}}
\newcommand{\eeas}{\end{eqnarray*}}
\newcommand{\ba}{\begin{array}}
\newcommand{\ea}{\end{array}}
\newtheorem{th}{Theorem}
\newcommand{\tl}{\tilde}
\newcommand{\rar}{\rightarrow}
\newcommand{\lar}{\leftarrow}
\newcommand{\lrar}{\longrightarrow}
\newcommand{\llar}{\longleftarrow}
\newcommand{\fr}{\frac}
\newcommand{\pa}{\partial}
\newcommand{\mb}{\mbox}
\newcommand{\lft}{\lefteqn}
\newcommand{\hs}{\hspace}
\newcommand{\vs}{\vspace}
\newcommand{\hst}{\hspace*}
\newcommand{\vst}{\vspace*}
\newcommand{\lb}{\label}
\newcommand{\nl}{\newline}
\newcommand{\np}{\newpage}
\newcommand{\om}{\omega}
\newcommand{\Om}{\Omega}
\newcommand{\al}{\alpha}
\newcommand{\bt}{\beta}
\newcommand{\eps}{\epsilon}
\newcommand{\veps}{\varepsilon}
\newcommand{\ld}{\lambda}
\newcommand{\Ld}{\Lambda}
\newcommand{\gm}{\gamma}
\newcommand{\Gm}{\Gamma}
\newcommand{\sg}{\sigma}
\newcommand{\bib}{\bibitem}
\newcommand{\ct}{\cite}
\newcommand{\rf}{\ref}
\title{Bloch electron in a  magnetic
field : diagonalization of tight-binding models}
\author{\sc Alexander Moroz\dag}
\date{\it Institute of Physics \v{C}SAV\\
\it  Na Slovance 2 \\
\it CS-180 40 Prague 8, Czechoslovakia}
\maketitle
\begin{center}
{\large\sc abstract}
\end{center}
A connection of a variety of tight-binding models
of noninteracting electrons on a rectangular lattice
in a magnetic field with theta functions is established.
A new  spectrum generating symmetry is discovered which
essentialy reduces the problem of diagonalization of these models.
Provided that one knows one eigenvector at one point in the parameter
space of
the corresponding
Harper equation
one knows an eigenfunction of
the corresponding model in the whole range of momentum
singlet out by the Landau gauge.
\maketitle
\newpage
The classical problem of Bloch electrons in magnetic fields
has been studied by
many authors \cite{Ha}, however, despite some general rigorous
results about the spectrum of the problem \cite{Ha,BS}
the analytic expressions for
the energy spectrum and wave functions are still unknown.
In what follows a whole variety of tight-binding ($t-b$) models
of noninteracting
electrons in a magnetic field is discussed.
We shall start with a rational magnetic field, $\al =p/q$,
describe symmetries of these  models and
establish a
connection  with theta functions with chracteristics \ct{M}.
In our construction the whole symmetry of the Hamiltonian of $t-b$
models is used which is the invariance under continuos
magnetic translations (MT's), i.e., under an action of the
full Heisenberg group - a fact which has been ignored in previous
discussions of the problem \ct{Ha,BS}.
The use of theta functions is then natural since they
arise in connection with representations of finite dimensional
subgroups of the Heisenberg group \ct{M} which are
relevant for a description of the spectrum of the Bloch electron.
The ordinary theta functions have appeared in the study of a
related problem \ct{AB},
however, so far as we know nobody applied theta functions with
characteristics to the study of the spectrum of
$t-b$ models \ct{Mo}. The problem of diagonalization of these models
or the problem of solving
a polynomial equation \ct{Ha} is then reduced to one point
of the Brillouin zone (BZ) for the momentum singlet out by the Landau
gauge.
This is shown by using {\em modular transformations}, $SL(2,Z)$ \ct{M}.

Let us start with some definitions. In what follows
the Landau gauge \(\vec{A}=B(-y,0,0)\) and a substrate
potential periodic under translations by
$\vec{a_1}$ and $\vec{a_2}$
 in $x$ and $y$ direction, respectively, are assumed.
A truncation of the Fourier cosine expansion of the
substrate potential accomapanied with the
Peierls substitution then leads on the Hamiltonian of a
$t-b$ model \ct{Ha} including in general terms describing
$j$-th-nearest-neighbour hopping up to $j\leq n$,
\begin{equation}
{\cal H} = t_1 (S_{\vec{a_1}} + S_{\vec{a_1}}^* + S_{\vec{a_2}} +
S_{\vec{a_2}}^* ) + \ldots,
\label{Ham}
\end{equation}
where $S_{\pm\vec{a_\ell}}$ ($S_{-\vec{a_\ell}}=S^*_{\vec{a_\ell}}$)
is a shift operator,
$S_{\pm\vec{a_j}} = e^{\pm\frac{i}{\hbar}a_j m\hat{v}_j}$,
$\hat{v}_j$, $j=1,2$, being components of
the standard velocity operator, and $\ldots$ in (\rf{Ham}) stands
for integer powers $j$ of the shift opertors multiplied by a
corresponding overlap integral $t_j$ and describing
next-nearest-neighbour hopping, etc.
Action of $S_{\pm\vec{a_j}}$ on a function $\psi (\vec{r})$ is given
as follows,
\begin{equation}
\begin{array}{rcccl}
S_{\pm\vec{a_1}} \psi (\vec{r}) & = & e^{\pm\frac{i}{\hbar}
a_1(\hat{p}_1
+\frac{e}{c} By)} \psi (\vec{r}) & = & \psi (\vec{r}\pm\vec{a_1})
e^{\pm 2\pi i\alpha y/a_2}, \nonumber\\
S_{\pm\vec{a_2}} \psi (\vec{r}) & = & e^{\pm\frac{i}{\hbar}
a_2\hat{p}_2}
\psi (\vec{r}) & = & \psi (\vec{r}\pm\vec{a_2}) ,\label{ht}
\end{array}
\end{equation}
where according to Ref. \cite{Ha} $\alpha=\Phi /\Phi _o$, $\Phi$ and
$\Phi _o=hc/e $ being  the magnetic flux through an elementary
plaquette and the flux quantum, respectively.
Although recently a paper appeared where the
use of the Peierls substitution is shown to lead to incorrect
conclusions for total energy of electrons in a magnetic field \cite{Al},
single electron energies are not affected (after some shift and
rescaling) by this substitution \cite{Al1}.
The displayed part of the hamiltonian ${\cal H}$ which is the
Hamiltonian of the nearest-neighbour $t-b$ model leads
on the so-called Harper equation (HE) \ct{Ha},
\be
({\cal H}u)(n) = u(n+1) + u(n-1) + 2 \cos (2\pi\al n
+\Theta)u(n) = E u(n).
\lb{ha}
\ee
Here  $u(n) = g(n)$, where $g(n)$ enters a parametrization of the wave
function:
$\psi (ma_1, na_2)=e^{ik_1 m}g(n)$, $k_1$ being component of the Bloch
momentum, and $\Theta =k_1 a_1$.

Commutation relations for the components
of the velocity operator in a magnetic field imply the
following commutation relations for the shift operators
$S_{\pm\vec{a_j}}$, $j=1,2$,
\be
S_{\vec{a_1}} S_{\vec{a_2}} = S_{\vec{a_2}}S_{\vec{a_1}}e^{-2\pi i\al}.
\lb{s-cr}
\ee
By a simple consideration one finds that $1/\al$ roots (powers) of
the shift operators, i.e., the Azbel operators $\hat{A}$ and $\hat{B}$ \ct{Ha},
$\hat{A} := S_{\vec{a_1}}^{1/\al}$,
$\hat{B} := S_{\vec{a_2}}^{1/\al}$,
commute with the Hamiltonian ${\cal H}$. In contrast to the shift
operators their commuation relation is
\be
\hat{A}\hat{B} = \hat{B}\hat{A} e^{2\pi i/\al}.
\lb{c-ab}
\ee
The above defined operators are not the only
ones which commute with ${\cal H}$. One can choose
for generators of operators commuting with ${\cal H}$
components of the velocity operator transformed under spatial parity.
Indeed working in the above Landau gauge one can check that any of
the two
operators from the set $\{\hat{p_1}+\frac{e}{c}By,\hat{p_2}\}$
commutes with both operators from the
set  $\{\hat{p}_1,\ \hat{p}_2 + \frac{e}{c}Bx\}$
and vice versa. Hence, the operators
$T_{\pm\xi\vec{a_j}}$,
$j =1,2$, called operators of  magnetic translations (MTO),
\begin{equation}
T_{\pm\xi\vec{a_1}} := e^{\pm\xi\frac{i}{\hbar}a_1 \hat{p}_1},\hspace{2cm}
T_{\pm\xi\vec{a_2}} := e^{\pm\xi(\frac{i}{\hbar}a_2 \hat{p}_2 + 2\pi
i\alpha x/a_1)},
\label{mt}
\end{equation}
commute with ${\cal H}$ for any $\xi$.
One can justify that the spatial parity transformed components
of the velocity operator are not gauge equivalent to the primary ones
(the model under consideration is  not (except for $\al =1/2$)
parity invariant),
and hence MTO's are independent of
the shift operators, as well as of $\hat{A}$ and $\hat{B}$.
Whenever $\xi$ is an integer we shall call MTO's as the lattice
or integer MTO's.
The lattice MTO's form a projective (ray) representation of the translation
group. For any $\alpha$:
\be
T_{\vec{a_1}}T_{\vec{a_2}}=T_{\vec{a_2}}T_{\vec{a_1}}
e^{-2\pi i\alpha},\hspace{2cm} T_{\vec{a_1}}T_{\vec{a_2}} =
T_{\vec{a_1}+\vec{a_2}} e^{i\pi\alpha}.
\lb{c-mt}
\ee
Therefore, in virtue of (\rf{c-ab},\rf{c-mt}),
as a maximal set of commuting operators can be taken any set of the
form
\be
\{ {\cal H}, \hat{A}^{\rho_1}, \hat{B}^{\rho_2},
T_{\sg_1\vec{a_1}}, T_{\sg_2\vec{a_2}} \mid
\rho_j \in Z,\ \rho_1\rho_2 \sim \al;\ \sigma_j \in R,
\sigma_1 \sigma_2 \sim 1/\al;\ j=1,2 \}.
\lb{cs}
\ee
In a rational magnetic
field $\al =p/q$, $p$ and $q$ being relative prime integers, one
can always set $\rho_1=1,\ \rho_2=p,\ \sigma_1=1,\ \sigma_2=q$.
Note that the ``minimal''commuting MTO's  are
$T_{\vec{a_1}}$ and $T_{q\vec{a_2}/p}$ in this case. However,
$T_{q\vec{a_2}/p}$
doesn't relate points which are related by ${\cal H}$ and hence
is unimportant for the classification of the spectrum of ${\cal H}$.
As a whole $t-b$ models have more symmetry
than the model described by the Hamiltonian
$ H = (1/2m)(\vec{p}-\frac{e}{c} \vec{A})^2 + V(\vec{r})$,
with a general substrate potential $V(x,y)$ periodic under
translations by $\vec{a_j}$.
${\cal H}$ is invariant under continuos MT's (if considered in a
suitable
Hilbert space (see below))
while $H$ only commutes with the lattice MTO's
\ct{L} which imply the following
periodicity conditions for an eigenfunction $\psi (\vec{r})$
of $H$,
\be
\psi (x\pm a_1,y) = \psi '(x,y),\hs{1cm}
\psi (x,y\pm a_2)  =  \psi '' (x,y)e^{\mp 2\pi i\alpha x/a_1},
\lb{per}
\ee
$\psi '$ and $\psi ''$ being from the same degenerate set
as $\psi$.

Moreover, apart continuos MTO's, $H$ does not
commute with the operators $\hat{B}$ and
$\hat{A}$.
Therefore the procedure which leads on the
$t-b$ Hamiltonian ${\cal H}$ give a more symmetric Hamiltonian
than the primary one. This is why, as we shall see, the problem of
finding the spectrum of
${\cal H}$ for a given crystal momentum $\vec{k}$ and a given
magnetic field can be essentially reduced.
However, a crucial difference between  finite-order
differential equation and finite-difference equation
(i.e., infinite-order differential one) should be noted.
The additional symmetries of ${\cal H}$ with regard to $H$ don't
lead to any additional degeneracy of energy levels since
they don't
relate points which are related by the Hamiltonian.
The additional symmetries result in that we have a continuous family
of equivalent Hilbert spaces associated with a fixed lattice.
They differ each other by the boundary
conditions imposed (see below; for a similar case see \ct{FH}).
Therefore, in a rational magnetic field $\al =p/q$
the spectrum of either ${\cal H}$ or $H$ can be
classified by irreducible representations of the
magnetic group or by a discrete subgroup of the Heisenberg group
generated by $\{T_{\vec{a_1}}, T_{q\vec{a_2}}\}$,
respectively \ct{B}.
Action of $T_{\vec{a_2}}$ which is not in a commuting
set (\rf{cs}) then leads to a  $q$-fold degeneracy of energy levels
which implies the well-known Diophantine equation for the
Hall conductance $\sigma $ \ct{D},
$p\sigma + qm = 1$, $m$ being an integer.
Irreducible representations of
the magnetic subgroup are one dimensional and
can be classified by quasi-momenta $\vec{k}$ which take on
the values in the first magnetic BZ (MBZ).
Since the
spectrum of either $H$ or ${\cal H}$ is periodic under translations
$\alpha\rightarrow\alpha\pm 1$, it will be assumed that
$\alpha\in [0,1)$.

The Heisenberg group $G$, which is nothing but the central extension
of the group of ordinary lattice translations, can be defined as a
set of elements
$\{(\lambda,a,b) | \lambda\in C^*; a,b \in R\}$ with the
multiplication law:
$(\lambda ,a,b)(\lambda ', a',b')=(\lambda\lambda '
e^{-2\pi i (ba'-b'a)},a+a',b+b'),$
where $C^*$ is a unit circle in the complex plane. Let us denote
by $q\Gamma$ a discrete subgroup of $G$,
$q\Gamma :=\{(1,qa,b) | a,b \in Z \}$.
We shall denote by $V$ the Hilbert space of entire functions
$f(z)$ with the norm induced by the scalar product,
\be
||f||^2 = \int \exp (-2\pi y^2 /Im\tau )|f(x+iy)|^2\  dx\ dy,
\lb{norm}
\ee
where the integral is taken over the elementary periodicity domain,
$\tau$ being a modular parameter (modulus).
Let $V_q$ be a  subspace of $V$
invariant under $q\Gamma$. The action of
$(1,qa,b)\in q\Gamma$ on $f(z)\in V_q$ is given as usual,
$(1,qa,b)f(z)= e^{\pi ib^2\tau +2\pi ib(z+qa)}f(z+qa+b\tau),$
and $f(z)\in V_q$ if and only if
$f(z) = \sum _{n\in (1/q)Z} c_n \nl\exp \{\pi in^2\tau + 2\pi inz\},$
with $c_{n+1} =c_n$, i.e., $V_q$ is a $q-$dimensional (complex)
subspace of $V$.
Then a discrete subgroup $G_q$,
$G_q := \{(\lambda,a,b) |\lambda\in C^*_q, a\in Z, b\in (1/q)Z\}
/$(mod $q\Gamma$) = $C^*_q\times Z/qZ \times (1/q)Z/Z $,
$C^*_q$ being the cyclic group of $q-$roots of 1, commutes with
$q\Gamma$. Following readily arguments in \cite{M} step by step
one can show that the finite
group $G_q$ acts irreducibly on $V_q$. Moreover, one has even
an analoque of the Stone-von Neumann theorem for discrete subgroups
of the Heisenberg group \cite{M} .
Because of irreducibility the action of $G_q$ on $V_q$ determines a
canonical basis for $V_q$ and $G_q$ acts in a fixed way. The standart
basis of $V_q$ is given in terms of theta functions with
upper characteristic $\ell\alpha$ (modulo a constant), where
$\ell = 0,1,\ldots ,q-1$ \ct{M}.
For a given modulus $\tau$ let the complex
variable $z$ be defined by $z= z(\tau):= \tau\al y/a_2$ and
let us consider $x/a_1$ as the lower characteristic of the
modified theta function
\( g[\stackrel{\ell\al}{\scriptstyle x/a_1}]
(\tau |\tau\al y/a_2) \) defined
as follows,
\be
g [\stackrel{\ell\al}{\scriptstyle x/a_1}]
(\tau |\tau\al y/a_2) :=
e^{\pi i\tau y^2\al ^2/a_2^2}\theta
[\stackrel{\ell\al}{\scriptstyle x/a_1}](\tau |z(\tau)).
\lb{bf}
\ee
$\theta [\stackrel{\ell\al}{\scriptstyle x/a_1}](\tau |\tau\al\ y)\in V_\al$
is the usual Jacobi theta function with characteristics
\ct{M},
\begin{equation}
\theta [\stackrel{\ell\al}{\scriptstyle x/a_1}]
(\tau |\tau\al y/a_2) = \sum_{n=-\infty}^{\infty}
\exp\{\pi i\tau (n+\ell\alpha )^2 + 2\pi i(n+\ell\alpha )
(\frac{x}{a_1}+\tau \alpha\frac{y}{a_2})\}.
\label{theta}
\end{equation}

In what follows we shall rescale our coordinates according to:
$x_j/a_j\rightarrow x_j$.
In the following step we shall introduce the momenta $(k_1,k_2)$ and
define functions
$g^{k_1k_2}
[\stackrel{\ell\al}{\scriptstyle x}]
(\tau|\tau\al y) := e^{ik_1x +ik_2y}g[\stackrel{\ell\al}{\scriptstyle x}]
(\tau|\tau\al y)$.
Let us denote the Hilbert space of functions
$g^{k_1k_2} [\stackrel{\ell\al}{\scriptstyle x}]$
with the scalar product (\ref{norm})
by $W_\al$.
For a given modulus $\tau$ the lattice MTO's (considered
as a subgroup of
the Heisenberg group) provide the
unitary irreducible
representation (UIR) of $G_q$ in $W_\al$,
$G_q \ni (1,\pm 1,0)
\hookrightarrow T_{\pm\vec{a_1}}$,
$G_q\ni (1,0,\pm\alpha )
\hookrightarrow T_{\pm\vec{a_2}}$.
Note that in a rational magnetic field, $\al =p/q$, irreducible action
of the lattice MTO's  requires that
the Hilbert
space be specified by periodic boundary conditions (PBC) on the square
defined by translations by $q$ {\em lattice spacings in both directions}
despite that already $T_{\vec{a_1}}$ and
$T_{q\vec{a_2}}$ commute (\rf{c-mt}).
The condition that a state $\psi (x,y):= \sum_\ell  d_\ell
g^{k_1k_2}[\stackrel{\ell\al}
{\scriptstyle x}](\tau |\tau\al y)$ is an eigenfunction
of the  rescaled  Hamiltonian ${\cal H}$,
${\cal H}= {\cal H}/t_1$ at the point $(x_o,y_o)$ of the plane is
tantamount
to the condition that the $q$-dimensional vector $\vec{d}=(d_o,d_1,\ldots
,d_{q-1})$
solves the HE,
\be
e^{-iK_2} d_{\ell +1} +
e^{iK_2}
d_{\ell -1} + 2\cos [(K_1 +2\pi\ell\al ] d_\ell =\varepsilon
d_\ell ,
\lb{h2}
\ee
at $K_1 = k_1 +2\pi\al y_o$ and $K_2 = k_2 -2\pi\al x_o$, where
$\varepsilon =E/t_1$. In the matrix notation the vector $\vec{d}$ has to
be the eigenvector of the following hermitian $q\times q$ matrix:
\begin{equation}
\left( \begin{array}{lccc}
C_o & e^{-iK_2 } &  \ldots & e^{iK_2}   \\
e^{iK_2 } & C_1 & \ldots & 0 \\
0  & e^{iK_2 } & \ldots & 0   \\
\vdots & \vdots &  \vdots &\vdots                   \\
0      &0 & \ldots  &       e^{-iK_2}                 \\
e^{-iK_2} & \ldots &e^{iK_2} & C_{q-1}
\end{array}  \right) ,
\label{m}
\end{equation}
where $C_\ell$ stands for $2\cos (K_1
+ 2\pi\ell\al)$.
To show this one can use the behaviour of $g[\stackrel{\ell\al}
{\scriptstyle x}](\tau |\tau\al y)$ under lattice
translations,
\begin{eqnarray}
g[\stackrel{\ell\al}{\scriptstyle x\pm 1}]
(\tau |\tau\al y) = e^{\pm 2\pi i\ell\al}g[\stackrel{\ell\al}{\scriptstyle x}]
(\tau |\tau\al y) ,
\nonumber \\
g[\stackrel{\ell\al}{\scriptstyle x}]
(\tau |\tau\al(y\pm 1)) =
e^{\mp 2\pi i\al x}g[\stackrel{(\ell\pm 1)\al}{\scriptstyle x}]
(\tau |\tau\al y).
\lb{pc}
\end{eqnarray}
Note that for a given
parameters  $(K_1,K_2)$ of the HE
we have a {\em freedom} to choose $(x_o,y_o)$
{\em at our convenience}
since a change in $(x_o,y_o)$ can be compensated by adjusting the values
of the momenta
$(k_1,k_2)$. Due to the periodicity properties (\ref{pc}) of theta
functions the same HE (for a given momenta $(k_1,k_2)$)
(\ref{h2}) will be repeated at the
points of the lattice $(x_o +qm,y_o +nq)$, $m$ and $n$ being integers.
Since, for a given $(x_o,y_o)$,
\bea
g[\stackrel{\ell\al}{\scriptstyle x_o +m  \pm q }]
(\tau |\tau\al (y_o +n)) =
g[\stackrel{\ell\al}{\scriptstyle x_o +m}]
(\tau |\tau\al (y_o+n)),
\nonumber\\
g[\stackrel{\ell\al}{\scriptstyle x_o +m}]
(\tau |\tau\al (y_o +n \pm q))
= e^{\mp iq2\pi\al x_o}
g[\stackrel{\ell\al}{\scriptstyle x_o +m}]
(\tau |\tau\al (y_o+n )),
\lb{xpc}
\eea
basis functions
$g^{k_1k_2}[\stackrel{\ell\al}{\scriptstyle x}]
(\tau |\tau\al y)$ satisfy {\em the same boundary conditions} (BC) on the
elementary periodicity $q\times q$ square at any point of
the $(x_o +m,y_o +n)$ lattice.
This corresponds to the fact that
these points are connected by the Hamiltonian ${\cal H}$ and states
on this lattice belong to {\em the same} Hilbert space.
Because of (\ref{xpc})
any basis function carries an internal Bloch momentum
$k_2=-2\pi\al x_oy$ on the $(x_o,y_o)$
lattice.  Thus, the state $\psi (x,y):= \sum_\ell  d_\ell
g^{k_1k_2}[\stackrel{\ell\al}
{\scriptstyle x}](\tau |\tau\al y)$ carries the Bloch momenta
$(k_1,K_2)$ on the $(x_o,y_o)$ lattice.
The Hilbert space on a given $(x_o,y_o)$ lattice is determined by
the scalar product
\bea
\lefteqn{\langle g^{k_1k_2}[\stackrel{s\al}{\scriptstyle x}]
|g^{k_1'k_2'} [\stackrel{\ell\al}{\scriptstyle
x}]\rangle_{W_\al(x_o,y_o)}:=}
\nonumber\\
&&
\sum_{0\leq m,n\leq q-1}
\overline{g^{k_1k_2}[\stackrel{s\al}{\scriptstyle x_o+m}]
(\tau |\tau\al y_o+n)} g^{k_1'k_2'}[\stackrel{\ell\al}{\scriptstyle
x_o+m}] (\tau|\tau\al y_o+n).
\lb{spr}
\eea
One can check that
$W_\al (x_o,y_o)$ is the Hilbert space, ${\cal H}$ is
the hermitian operator, and that
$g^{k_1k_2}[\stackrel{\ell\al}{\scriptstyle x}]{}'s$
are
orthogonal for a given $(k_1,k_2)$,
$\langle .\ [\stackrel{s\al}{\scriptstyle x}]
(\tau |\tau\al y)
|.\ [\stackrel{\ell\al}{\scriptstyle x}]
(\tau |\tau\al y)\rangle_{W_\al(x_o,y_o)}=
q\delta_{s\ell}\sum_{n=o}^{q-1}
\overline{g[\stackrel{n\al}{\scriptstyle x_o}](\tau |\tau\al y_o)}
g[\stackrel{n\al}{\scriptstyle x_o}](\tau |\tau\al y_o)$
The notion of eigenfunction on a given lattice is selfconsistent
since if
$\psi (x,y)= \sum_\ell d_\ell(K_1,K_2) g^{k_1k_2}\nl [\stackrel{\ell\al}
{\scriptstyle x}](\tau |\tau\al y)$ is the eigenfunction of ${\cal H}$
at a given point of the $(x_o,y_o)$ lattice
that it {\em remains} to be eigenfunction
of ${\cal H}$ for all points of
$(x_o,y_o)$ {\em lattice}. Indeed, at the point
$(x_o,y_o+1)$ one obtains the HE at $(K_1+2\pi\al,K_2)$.
By using the
periodicity properties (\ref{pc}) of the basis functions
one can pull back the equation to the original point $(x_o,y_o)$.
The above statement will be then proved if one
shows that $\psi (x,y)= \sum_\ell e^{-2\pi i\al x} d_{\ell-1}(K_1,K_2)
g^{k_1k_2}[\stackrel{\ell\al}
{\scriptstyle x}](\tau |\tau\al y)$ solves the HE at
the above parameters at the original point $(x_o,y_o)$. However, this
can be checked
rather straightforwardly.
Similarly, at the point $(x_o+1,y_o)$ one has to show that
$\psi (x,y)= \sum_\ell d_\ell(K_1,K_2) g^{k_1k_2}[\stackrel{\ell\al}
{\scriptstyle x}](\tau |\tau\al y)$ solves the HE at
$(K_1,K_2-2\pi\al)$. This is equivalent to show that
$\psi (x,y)= \sum_\ell e^{2\pi i\ell\al}d_\ell(K_1,K_2)
g^{k_1k_2}[\stackrel{\ell\al}
{\scriptstyle x}](\tau |\tau\al y)$ solves the HE at the
above parameters at the original point
$(x_o,y_o)$. This can be again justified rather straightforwardly.
Therefore any solution $d_\ell (K_1,K_2)$ of the HE defines
eigenfunction $\psi (x,y)= \sum_\ell d_\ell(K_1,K_2)
g^{k_1k_2}[\stackrel{\ell\al}
{\scriptstyle x}](\tau |\tau\al y)$ of ${\cal H}$ in the whole
continuous family of the Hilbert spaces of states on the
$(x_o,y_o)$ lattices if momenta $(k_1,k_2)$ and
the lattice $(x_o,y_o)$ are adjusted such that
the parameters $(K_1,K_2)$ are kept
fixed mod $2\pi\al$.
The continuous
MTO's can be viewed as operators which map states from one Hilbert
space of states associated with a given lattice to another.
By means of them one can put the eigenfunctions of ${\cal H}$
for different momenta $(k_1,k_2)$ and, of course, for
different $(x_o,y_o)$ lattices, {\em on a given lattice}, i.e.,
to {\em the same Hilbert space}.
Let us consider a state
$\psi (x,y) = \sum_\ell\ d_\ell g^{k_1k_2}[\stackrel{\ell\al}
{\scriptstyle x}](\tau |\tau\al y)
\in W_\al (x_o,y_o)$,
i.e., the $(x_o,y_o)$-lattice section of $W_\al$,
with the components $d_\ell = d_\ell (K_1,K_2)$,
which solves (\rf{h2}) for $(K_1,K_2)$ at the point $(x_o,y_o)$.
Then $\psi' := T_{\eps_j\vec{a_j}}\psi \in W_\al(\vec{r}_o
+\eps_j\vec{a}_j)$,
$j=1,2$, satisfies a different BC and it is  an  element of
$W_\al(\vec{r}_o +\eps_j \vec{a}_j)$. It
ceases to be an eigenvector since its components
$d_\ell$
have to satisfy the HE at
$(K_1+2\pi\al\eps_2, K_2-2\pi\al\eps_1)$, i.e.,
\be
e^{-i(K_2-2\pi\al\eps_1)} d_{\ell +1} +
e^{i(K_2-2\pi\al\eps_1)}
d_{\ell -1} + 2\cos (K_1+2\pi\ell\al) d_\ell =\varepsilon
d_\ell .
\lb{h3}
\ee
Thus, if we wish to
make the property that some state $\psi$ is an eigenfunction
{\em invariant under continuous MT's}, $T_{\eps_j\vec{a_j}}$, we have to
simultaneously transform either
the components $d_\ell (K_1,K_2)$ according to
\(d_\ell (K_1,K_2)\hookrightarrow d_\ell (K_1+2\pi\al
\eps_2,K_2-2\pi\al\eps_1)\),
or the momenta $(k_1,k_2)$ by the following rule:
$(k_1,k_2)\hookrightarrow(k_1-2\pi\al
\eps_2, k_2+2\pi\al\eps_1)$.
First, let us consider the combined
transformation (the shift by $T_{\eps_j\vec{a_j}}$ and the
transformation of
$d_\ell
(K_1,K_2)$) by
$U(\eps_1)$
and $U(\eps_2)$ separately. Therefore the MTO's don't commute
in general the image of $\prod_{j_s} U(\eps_{j_s})$ ($s$ being an
index, one or two,
which distinguishes between translations in $x$ and $y$ directions)
is only defined modulo an overall phase factor. To get rid of it
we shall define the operator $Q(\eps_1,\eps_2)$,
\be
Q(\eps_1,\eps_2) := [e^{i\vec{k}\vec{r}}(\prod_{j_s} T_{\xi_{j_s}a_s})
e^{-i\vec{k}\vec{r}}] \prod_{j_s} U(\xi_{j_s}),
\lb{qop}
\ee
where the both products are ordered in the same manner and $\eps_s =
\sum_{j_s}\xi_{j_s}$. From the above it follows that $Q(\eps_1,\eps_2)$
defines a periodic function
on the $(m,n)$-lattice, $m$ and $n$ being integers.
Similarly we shall
define the other combined mapping $R(\eps_1,\eps_2)$
(shift by $T_{\eps_j\vec{a_j}}$ and the transformation of
momenta $(k_1,k_2)$) by $R(\eps_1,\eps_2)$.
The operators $Q(\eps_1,\eps_2)$ and $R(\eps_1,\eps_2)$ defines
canonical mappings
between different Hilbert spaces $W_\al (x_o,y_o)$ under which
{\em eigenvectors are mapped on eigenvectors}.
Thus we know that for any $\eps_1$ and $\eps_2$ both
$Q(\eps_1,\eps_2)\psi_{(x_o,y_o)}:=\sum_\ell d_\ell
(K_1+2\pi\al\eps_2,K_2-2\pi\al\eps_1)
g^{k_1+2\pi\al\eps_2,k_2}[\stackrel{\ell\al}
{\scriptstyle x+\eps_1}](\tau |\tau\al (y+\eps_2))$
and
$R(\eps_1,\eps_2)\psi_{(x_o,y_o)}
:=\sum_\ell d_\ell (K_1,K_2)
g^{k_1k_2+2\pi\al\eps_1}[\stackrel{\ell\al}
{\scriptstyle x+\eps_1}](\tau |\tau\al (y+\eps_2))$
are eigenfunctions of ${\cal H}$ at $(x_o,y_o)$ if
$\psi_{(x_o,y_o)}:=\sum_\ell d_\ell
(K_1,K_2)
g^{k_1k_2}[\stackrel{\ell\al}
{\scriptstyle x}](\tau |\tau\al y)$ does so.
By using the {\em modular transformations}
we can
explicitely write
the induced change of the components $d_\ell$ down.
Let us suppose that
there exists
a relation between our basis functions at different points $(x,y)$ and
$(\dot{x},\dot{y})$, say
\be
g^{k_1k_2}[\stackrel{\ell\al}
{\scriptstyle x}](\tau |\tau\al y) = {\cal M}_{\ell s}(\tau
|x,y)g^{k_1k_2}[\stackrel{\ell\al}
{\scriptstyle \dot{x}}](\dot{\tau} |\dot{\tau}\al \dot{y}).
\lb{modt}
\ee
Our main idea is as follows.
Let us assume that there exists a {\em parametrization} of the components
$d_\ell =F(\bar{d}_\ell)$
of a vector $\psi (x,y) = \sum_\ell d_\ell g^{k_1k_2}[\stackrel{\ell\al}
{\scriptstyle x}](\tau |\tau\al y)$ which exactly
cancels the effect of ${\cal M}_{\ell s}(\tau
|x,y)$ in the transformed basis under the action of ${\cal H}$. Then
the condition that $\psi$ is an eigenfunction on a given point
$(x_o,y_o)$ lattice is tantamount to the condition that both, $d_\ell$ and
$\bar{d}_\ell$, solve the HE. However, the equation
for $d_\ell$ is in general {\em at
different point of the BZ} than the equation for $\bar{d}_\ell$.
The parametrization then in turn induces a mapping between components
of
eigenvectors at these points of the BZ.
In what follows we shall look for relations between our basis
functions of the above type (\ref{modt}).
The simplest relations between the spectrum of ${\cal H}$
at different points of the BZ follow from the behaviour
of $g^{k_1k_2}[\stackrel{\ell\al}
{\scriptstyle x}]{}'s$ under lattice translations (\ref{pc}):
$d_\ell (K_1,K_2) = e^{2\pi i\ell\al}
d_\ell (K_1,K_2 -2\pi\al)$ and
$d_\ell (K_1,K_2)= e^{2\pi i\al x} d_{\ell -1}(K_1 +2\pi\al, K_2)$.
Since $g^{k_1k_2}[\stackrel{\ell\al}
{\scriptstyle x}](\tau |\tau\al (y\pm 1/\al))= e^{\mp 2\pi ix}
g^{k_1k_2}[\stackrel{\ell\al}
{\scriptstyle x}](\tau |\tau\al y)$
the action of $Q(0,\pm 1/\al)$ implies the relation $d_\ell (K_1,K_2)=
d_\ell (K_1\pm 2\pi,K_2)$.
Now we shall show that the required relation (\ref{modt}) is provided
by {\em modular transformations} \ct{M}.
The modular transformations are transformations of the form,
\be
\tau\hookrightarrow \dot{\tau} = \frac{d\tau +b}{c\tau +a},\hs{1cm}
z\hookrightarrow \dot{z} =
\frac{z}{c\tau +a} ,
\lb{mod}
\ee
the parameters $a,b,c,d$ being integers, $ad-bc=1$,
which form the group of modular transformations, $SL(2,Z)$.
If the Jacobi theta functions with chracteristics are
transformed under the modular transformation \ct{M} then
the transformation induces the following
identity between the modified theta functions:
\be
g^{k_1k_2}[\stackrel{\ell\al}{\scriptstyle x+b\al y}](\tau |\tau\al ay) =
(c\tau +d)^{-1/2}u_\ell e^{-i\pi ab\al^2 y^2}
g^{k_1k_2}[\stackrel{\dot{\ell}\al}{\scriptstyle \dot{x}}]
(\dot{\tau}|\dot{\tau}\al y),
\lb{modtr}
\ee
where $\dot{\ell}\al =\ell\al d -cx +cd/2$, $\dot{x}=
ax -b\ell\al +ab/2$, and $u_\ell$ is a phase factor which doesn't depend
on $\tau$ and $y$.
One of the key points in our
construction is that the (finite dimensional) Hilbert space $W_\al$
constructed above
{\em is not the unique choice}. By using the periodicity properties
of modified theta functions (\ref{pc})
one can show that the representation space
of the lattice MTO's (\ref{c-mt}) can be taken to be the Hilbert space
generated by
the modified theta functions from either right or left hand side of
Eq. (\ref{modtr}), as well.
Of course,
the parameters of the modular transformation (\ref{mod}) have to be such
that $(a-1)$, $b\al$, and $(d-1)$ are proportional to $q$.
Then, the contribution $-b\ell\al$ and $ab/2$ to $\dot{x}$ can be
ignored, the
upper characteristic remains rational (modulo a constant) and
proportional to $\al$, etc.
Moreover, one can
show that $u_\ell$ {\em doesn't depend} on $\ell$ in this case. One can also
check that these modular transformations are not empty. One of them is, e.g.,
$a=(\bar{a}q^2+1)$, $d=(\bar{d}q^2+1)$, $b=\bar{d}q^2$,
and $c=\bar{a}(q^2+k)$, whenever $\bar{a}+\bar{d}=k\bar{a}\bar{d}$.

Now, if one considers a vector
$\vec{d}=\sum_\ell d_\ell(K_1,K_2) g^{k_1k_2}[\stackrel{\ell\al}{\scriptstyle
x+b\al y}]
(\tau |\tau\al ay)$, then the condition that $\vec{d}$ is an
eigenvector of ${\cal H}$ at $(x_o,y_o)$ is tantamount to say that
the components  of $\vec{d}$ satisfy the Harper equation
at $(K_1,K_2)\in$ BZ, where $K_1=k_1+2\pi\al y_o$, and
$K_2=k_2-2\pi\al a(x_o+b\al y_o)$.
Now, if one {\em parametrizes}  the components
$d_\ell (K_1,K_2)$ of $\vec{d}$ according to
$d_\ell(K_1,K_2)=
e^{i\pi ab\al^2(y_o-\ell)^2}\bar{d}_\ell$ then one gets the
HE  for $\bar{d}_\ell$ at $(K_1,K_2+2\pi ab\al^2y_o)$.
The parametrization is taken to be such that it exactly cancels
the effect of $e^{-i\pi ab\al^2 y^2}$ if one looks
at the transformed basis (\ref{modtr}). Therefore
$d_\ell (K_1,K_2+2\pi ab\al^2y_o)=
e^{-i\pi ab\al^2 (y_o-\ell)^2}d_\ell(K_1,K_2)$.
Keeping $(K_1,K_2)$ fixed and changing $(k_1,k_2)$ and $(x_o,y_o)$
one can distribute the eigenvector at $(K_1,K_2)\in$ BZ throughout
the whole range of $K_2$.

By close inspection one finds that the same construction can be done
{\em for all
$t-b$ models discussed here} (including asymmetric ones
with direction dependendent overlap, $t^x_j\neq t^y_j$).
This is a rather straightforward consequence of the invariance
of ${\cal H}$ under the Heisenberg group.
More detailed consideration of the above symmetry with regard to
the diagonalization of $t-b$ models on a more general lattice and
a consideration of the flux phases will
be pursueded elsewhere.

I should like to thank D. Mumford
for sending me a copy of his book.
Discussions with
J. Bellissard, F. Bien, and D. J. Thouless on the related
problems, critical remarks by J. Fr\"{o}hlich, as well as
a financial support by
the Deutche Forschungsgemeinschaft in an early stage of
this work
and the present support by the Swiss National Foundation
are also gratefully acknowledged.
\vspace{0.3cm}

\noindent \dag Address after October 1, 1991:
Theoretische Physik, ETH-H\"{o}nggerberg, CH-8093 Z\"{u}rich,
Switzerland

\end{document}